\documentclass[a4paper,11pt]{article}
\usepackage{pos}

\usepackage{caption}
\usepackage{subcaption}

\title{Machine Learning Trivializing Maps: A First Step Towards Understanding How Flow-Based Samplers Scale Up}
\ShortTitle{Machine Learning Trivializing Maps}

\author{Luigi Del Debbio}
\author*{Joe Marsh Rossney}
\author{Michael Wilson}

\affiliation{Higgs Centre for Theoretical Physics, School of Physics and Astronomy, The University of\\ Edinburgh,
  Edinburgh EH9 3FD, UK}

\emailAdd{Luigi.Del.Debbio@ed.ac.uk, Joe.Marsh-Rossney@ed.ac.uk}

\abstract{
A trivializing map is a field transformation whose Jacobian determinant exactly cancels the interaction terms in the action, providing a representation of the theory in terms of a deterministic transformation of a distribution from which sampling is trivial.
Recently, a proof-of-principle study by Albergo, Kanwar and Shanahan [\href{https://journals.aps.org/prd/abstract/10.1103/PhysRevD.100.034515}{Phys. Rev. D \textbf{100} 034515 (2019)}] demonstrated that approximations of trivializing maps can be `machine-learned' by a class of invertible, differentiable neural models called \textit{normalizing flows}.
By ensuring that the Jacobian determinant can be computed efficiently, asymptotically exact sampling from the theory of interest can be performed by drawing samples from a simple distribution and passing them through the network.
From a theoretical perspective, this approach has the potential to become more efficient than traditional Markov Chain Monte Carlo sampling techniques, where autocorrelations severely diminish the sampling efficiency as one approaches the continuum limit.
A major caveat is that it is not yet understood how the size of models and the cost of training them is expected to scale.
As a first step, we have conducted an exploratory scaling study using two-dimensional $\phi^4$ with up to $20^2$ lattice sites.
Although the scope of our study is limited to a particular model architecture and training algorithm, initial results paint an interesting picture in which training costs grow very quickly indeed.
We describe a candidate explanation for the poor scaling, and outline our intentions to clarify the situation in future work.
}

\FullConference{%
 The 38th International Symposium on Lattice Field Theory, LATTICE2021
  26th-30th July, 2021
  Zoom/Gather@Massachusetts Institute of Technology
}

\begin{document}
\maketitle

\section{Sampling and Critical Slowing Down in Lattice Field Theory}

Numerical lattice field theory refers to a collection of strategies for approximating expectation values of the form
\begin{equation}
    \langle \mathcal{O} \rangle = \frac{1}{Z} \int \mathcal{D} \phi e^{-S(\phi)} \mathcal{O}(\phi) \, ,
    \qquad \mathcal{D}\phi = \prod_{x\in\Lambda} \mathrm{d}\phi_x \, ,
\end{equation}
by simulating field theories, defined on a space-time lattice $\Lambda \equiv a\mathbb{N}^d$ and described by an action $S(\phi)$, on a computer.
By simultaneously reducing the lattice spacing $a$ and keeping the physical volume fixed, one can obtain increasingly accurate extrapolations to the continuum limit with (in principle) quantifiable systematic errors.

This `simulation' is almost always a form of Markov Chain Monte Carlo (MCMC), whose function is to generate a representative sample of field configurations, i.e. one in which any configuration, $\phi$, may appear with a probability proportional to $e^{-S(\phi)}$.
Given such a sample of $N_\phi$ configurations, $\{\phi\}$, an unbiased estimate of $\langle \mathcal{O} \rangle$ is given by the sample mean,
\begin{equation} \label{eq:sample_mean}
    \overline{\mathcal{O}} = \frac{1}{N_\phi} \sum_{\{\phi\}} \mathcal{O}(\phi) \, , \qquad
    \mathrm{SE}_{\overline{\mathcal{O}}} = \sigma_{\mathcal{O}} \sqrt{\frac{2\tau_\mathcal{O}}{N_\phi}} \, .
\end{equation}

The statistical error on $\overline{\mathcal{O}}$, denoted $\mathrm{SE}_{\overline{\mathcal{O}}}$ above, depends on the process through which the sample was generated through the \textit{integrated autocorrelation time},
\begin{equation}
    \tau_\mathcal{O} = \frac{1}{2}\sum_{t=-\infty}^\infty \frac{\Gamma_\mathcal{O}(t)}{\Gamma_\mathcal{O}(0)} \geq \frac{1}{2} \, ,
\end{equation}
which is defined for a particular observable in terms of correlations between evaluations of that observable separated by $t$ steps in a thermalised Markov chain, $\Gamma_\mathcal{O}(t) = 
\left\langle \mathcal{O}(\phi^{(t)}) \mathcal{O}(\phi^{(0)}) \right\rangle - \langle \mathcal{O}\rangle^2$ (note that $\Gamma_\mathcal{O}(0) \equiv \sigma_\mathcal{O}^2$).
The integrated autocorrelation time is a measure of the statistical efficiency of an MCMC simulation; it takes approximately $2\tau_\mathcal{O}$ steps for the simulation to generate an effectively independent evaluation of $\mathcal{O}$.
An algorithm is said to suffer from \textit{critical slowing down} if $\tau_\mathcal{O}$ varies in proportion to the correlation length of the system (in units of $a$), $\xi$, raised to some power, $z_\mathcal{O} > 1$.
Unfortunately critical slowing down remains for the most part an intractable encumbrance in lattice QCD, where one would ideally like to perform continuum limit extrapolations with physical quark masses, implying large correlation lengths.

L\"{u}scher suggested a modification of the Hybrid Monte Carlo (HMC) algorithm~\cite{Duane1987} based on the idea of an invertible \textit{trivializing map} from the theory of interest to a limit in which the field variables decouple~\cite{Luscher2009}.
This original semi-analytical approach has yet to prove advantageous in practice.
However, the underlying idea --- that one can in principle construct a representation of the theory in which non-trivial aspects are encoded into a deterministic transformation of a distribution from which sampling is easy --- remains worthy of serious consideration.

Recently, this thread was picked up in a seminal study by Albergo, Kanwar and Shanahan~\cite{Albergo2019}, and followed up in Refs.~\cite{Kanwar2020,Boyda2020,Albergo2021b,Hackett2021}.
The key idea is that field transformations can be performed using an invertible neural network with a large number of adjustable parameters.
The task of constructing a trivializing map is cast as an optimisation problem; if the network is flexible enough, a careful tuning of the parameters (i.e. `training') can yield a good approximation of a trivializing map. 

\section{A Short Introduction to Normalizing Flows for Scalar Field Theory}

Consider the following action which describes a single-component scalar field with a quartic interaction term,
\begin{equation} \label{eq:action}
    S(\phi) = \frac{1}{2} \sum_{x\in\Lambda} \left[ -\beta \sum_{\mu=1}^d \phi_x \phi_{x+\hat\mu} + \phi_x^2 + \lambda(\phi_x^2 - 1)^2\right] \, .
\end{equation}
Let $\mathcal{F} : \mathbb{R}^{\lvert\Lambda\rvert} \to \mathbb{R}^{\lvert\Lambda\rvert}$ be a continuously differentiable bijective mapping whose effect is equivalent to taking $\beta,\lambda\to 0$, such that the probability density factorises into a product of univariate Gaussians,\footnote{In fact this definition may be overkill; the $\beta\to 0$ limit is not necessary to trivialize the theory.}
\begin{equation} \label{eq:trivializing_map}
    p\left( \mathcal{F}(\phi) \right)
    = p(\phi) \left\lvert \frac{\partial \mathcal{F}(\phi)}{\partial \phi} \right\rvert^{-1}
    = \mathcal{N} \prod_{x\in\Lambda} \exp\left(-\frac{\mathcal{F}(\phi)_x^2}{2} \right) \, ,
\end{equation}
where $p(\phi) \equiv Z^{-1} e^{-S(\phi)}$ and $\lvert \partial \mathcal{F}(\phi) / \partial \phi \rvert$ denotes the Jacobian determinant.
In this representation the degrees of freedom are decoupled Gaussian variables, which may be trivially sampled, while the correlated structure of $p(\phi)$ has been transferred to the Jacobian determinant of $\mathcal{F}$.
For convenience we will use $z \equiv \mathcal{F}(\phi)$ to label field configurations in this limit.

Our goal is to use neural networks to parametrise a transformation that, given sufficient training, inverts the trivializing map.\footnote{If we consider $\mathcal{F}$ to be an `encoder', we want to build the corresponding `decoder' that restores the information (correlations) we are interested in.}
Let this neural model, the \textit{normalizing flow}, be denoted $f_\theta$ where $\theta$ labels its adjustable parameters.
The training strategy proposed by Albergo, Kanwar and Shanahan~\cite{Albergo2019} is to first sample from $p(z)$, pass these Gaussian variates through $f_\theta$, and then finally adjust the parameters of the networks such that the following `loss function' decreases:
\begin{equation} \label{eq:loss_fn}
    L_\theta(\{z\}) = \frac{1}{N_z} \sum_{\{z\}} \left( S(f_\theta(z)) - \log \left\lvert \frac{\partial f_\theta(z)}{\partial z}\right\rvert \right) \, .
\end{equation}
Note that this involves computing the (logarithm of the) Jacobian determinant.
One hopes that iterating this training step many times will converge to an optimal set of parameters,
\begin{equation}
\theta^\star = \underset{\theta}{\mathrm{arg\,min}} \; L_\theta(\{z\}) \: \Rightarrow f_{\theta^\star} \: \approx \mathcal{F}^{-1} \, .
\end{equation}

Why should we expect this to work? Minimising the Eq.~\eqref{eq:loss_fn} with respect to $\theta$ is equivalent to minimising a stochastic estimator of the \textit{Kullbach-Leibler divergence}, 
\begin{equation} \label{eq:kullbach_leibler}
    D_\mathrm{KL}(\tilde{p} \Vert p) = \int \mathcal{D}\phi \tilde{p}(\phi) \log \frac{\tilde{p}(\phi)}{p(\phi)} \, ,
\end{equation}
which is a measure of distance\footnote{With a caveat being that it is asymmetric: $D_\mathrm{KL}(\tilde{p} ; p) \neq D_\mathrm{KL}(p ; \tilde{p})$.} between $p(\phi)$ and the transformed product of Gaussians,
\begin{equation}
    \tilde{p}(f_\theta(z)) = p(z) \left\lvert \frac{\partial f_\theta(z)}{\partial z} \right\rvert^{-1} \, .
\end{equation}
Using Eq.~\ref{eq:trivializing_map}, $D_\text{KL}(\tilde{p} \Vert p) = 0$ precisely when the Jacobian determinants of $f_\theta$ and $\mathcal{F}$ cancel. 

In general, $f_\theta$ must be a very flexible or `expressive' in order to invert $\mathcal{F}$, which we can safely assume is highly non-linear in the interesting case of a strongly interacting theory.
However, there is an important and opposing constraint which is that the Jacobian determinant must remain efficient to compute so that training does not become prohibitively expensive. 

\begin{figure}
\centering
    \includegraphics[width=.7\textwidth]{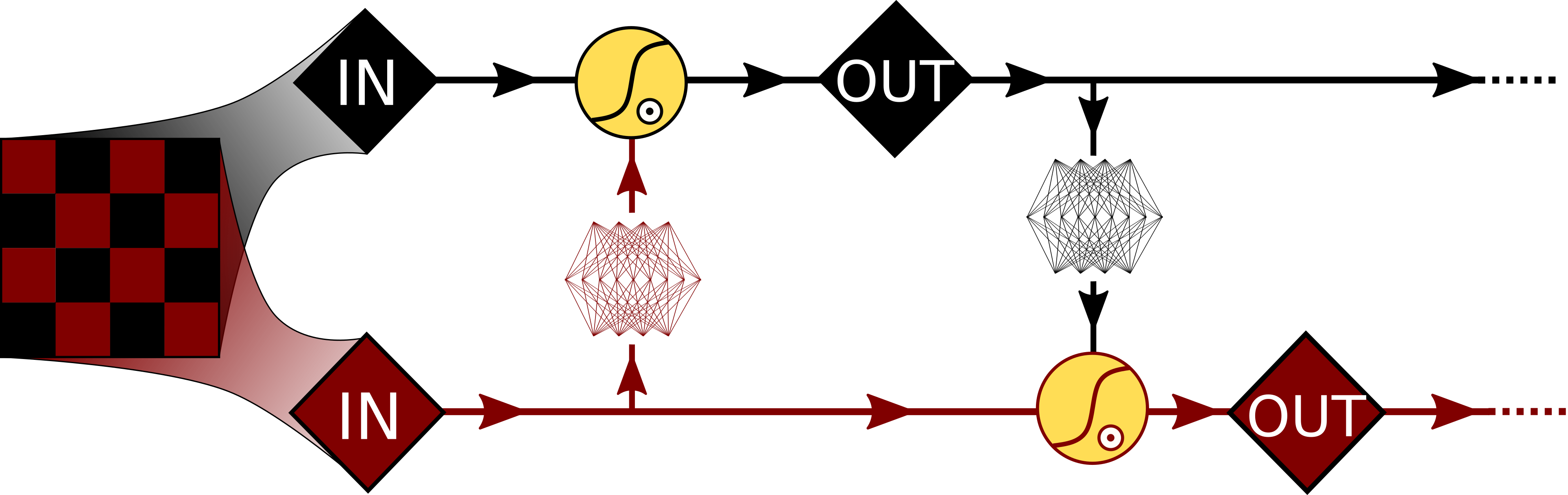}
    \caption{
    A visual representation of a pair of coupling layers using an even-odd `checkerboard' partitioning of the lattice, as described by Eq.~\eqref{eq:coupling_layer}. Let the inputs (`IN') represent a field configuration, $v_i$. Then the outputs (`OUT') represent $g_{i+1} \circ g_i \circ v_i = v_{i+2}$, where each element of $v_i$ has been transformed (exactly once) by a transformation that is parametrised by a complicated function of the elements from the opposite colour on the checkerboard. These functions are what the neural networks are trained to approximate.
    }
    \label{fig:coupling_layer}
\end{figure}

Though not the only approach, a popular compromise is to build the normalizing flow out of a sequence of transformations now widely referred to as \textit{coupling layers} \cite{Dinh2016}.
Coupling layers are essentially a template for building flexible, pointwise transformations that are guaranteed to have a triangular Jacobian matrix.
One divides the inputs to a coupling layer into two groups, only one of which will actually undergo a non-trivial transformation that is parametrised `on the fly' using information derived from the remaining, non-transformed inputs.
Considering a theory with a single degree of freedom at each lattice site, this implies a partitioning of the lattice into two disjoint subsets, $\Lambda^A$ and $\Lambda^P$, which we refer to as the `active' and `passive' partitions, respectively.
Motivated by locality, it is common to split the lattice into `even' and `odd' sites, and alternate between even-transforming layers and odd-transforming layers, as suggested in Figure~\ref{fig:coupling_layer}.

Defining a depth-$D$ normalizing flow as the composition of $D$ coupling layers, $f_\theta \equiv g_D \circ g_{D-1} \circ \ldots \circ g_2 \circ g_1$, and writing $v_1 \equiv z$ and $v_{D+1} \equiv \phi$, one can express the action of the $i$-th coupling layer as $g_i : v_i \mapsto v_{i+1} = g_i(v_i)$, where
\begin{align} \label{eq:coupling_layer}
    v_{i+1,x} = \begin{cases}
        v_{i,x} \, , &x\in\Lambda^P_i\\
        C_{i,x}\big(v_{i,x} \, ; \mathbf{N}_{i,x}(v_i^P ; \theta_i) \big) \, , &x\in\Lambda^A_i \, .
    \end{cases}
\end{align}
In the above, $v_i^P$ is short-hand for $\{v_{i, x} \mid x \in \Lambda^P_i\}$.
A set of invertible transformations $C_i$ (which are as yet unspecified) act on the active partition and depend on a set of parameters $\mathbf{N}_i(v_i^P; \theta_i)$ that are themselves functions of the passive variables.
The choice of labelling is suggestive of the fact that these functions are to be modelled by neural networks, with $\theta_i$ denoting the weights and biases.

The payoff for building $f_\theta$ out of these curious layers is that the logarithm of the Jacobian determinant reduces to nothing more than a sum over the log-gradients of every individual transformation in the normalizing flow, which is potentially extremely efficient to compute.
\begin{equation} \label{eq:log_det_jacob}
    \log \left\lvert \frac{\partial f_\theta}{\partial z} \right\rvert
    = \sum_{i=1}^{D} \sum_{x \in \Lambda_i^A} \log \left\lvert \frac{\partial C_{i, x}}{\partial v_{i, x}} \right\rvert \, .
\end{equation}
In fact this is the same Jacobian factor that would arise were we to replace the neural networks with parameters that had no dependence on the passive partition.
This is very nice because we can insert arbitrarily complex neural networks into the coupling layer at insignificant cost in the computation of $\log \tilde{p}(\phi)$. 

So far, we have described how one might construct a model that is capable of generating samples of \textit{independent} field configurations with probability proportional to $\tilde{p}(\phi)$, and how such a model may be trained so that $\tilde{p}(\phi) \approx p(\phi)$.
In practice, however, the quality of the approximation will almost certainly be insufficient to justify neglecting the biases that would arise by taking expectation values over $\tilde{p}(\phi)$ rather than $p(\phi)$.
Some form of reweighting is required to correct for these discrepancies.
Two approaches that have been used to date are, firstly, to weight the configurations generated by the model by running a Metropolis-Hastings simulation by using the model outputs as proposals that are accepted with a probability $A(\phi\to\phi^\prime)$ according to the `Metropolis test'~\cite{Albergo2019},
\begin{equation} \label{eq:metropolis_test}
    A(\phi \to \phi^\prime) = \min \left( 1, \frac{\tilde{p}(\phi)}{\tilde{p}(\phi^\prime)} \frac{p(\phi^\prime)}{p(\phi)} \right) \, ,
\end{equation}
and, secondly, to insert a reweighting factor $w(\phi) = p(\phi)/\tilde{p}(\phi)$ directly into Eq.~\eqref{eq:sample_mean}~\cite{Nicoli2020}.
Both of these approaches result in asymptotically exact sampling from $p(\phi)$, but crucially they rely on our ability to compute $\tilde{p}(\phi)$ exactly (up to a normalization) and efficiently.
In general, although one can construct training schemes that do not rely on an exact log-probability density for the model, it is essential for guaranteeing correct sampling.

The fraction of generated configurations that are accepted by the Metropolis test is a convenient metric for comparing models.\footnote{However, the acceptance rate is poor at diagnosing the issue of `mode-collapse': see Ref.~\cite{DelDebbio2021} Sec.VII/C and also Ref.~\cite{Hackett2021} Sec.VII/C.}
After verifying a monotonic correspondence with the integrated autocorrelation time, we generally used the acceptance fraction (or just `acceptance') in place of an estimate of $\tau_\mathcal{O}$ to quantify the sampling efficiency of trained models.

\section{Experiments and Scaling Study}

We have been experimenting with various recipes for building normalizing flows out of coupling layers, primarily with the aim of sampling from the action in Eq.~\eqref{eq:action}.
For greater detail on the transformations used, the experimental setup, and additional observations, please refer to Ref.~\cite{DelDebbio2021}.

\begin{figure}
\centering
    \begin{subfigure}{.49\textwidth}
        \includegraphics[width=.97\textwidth]{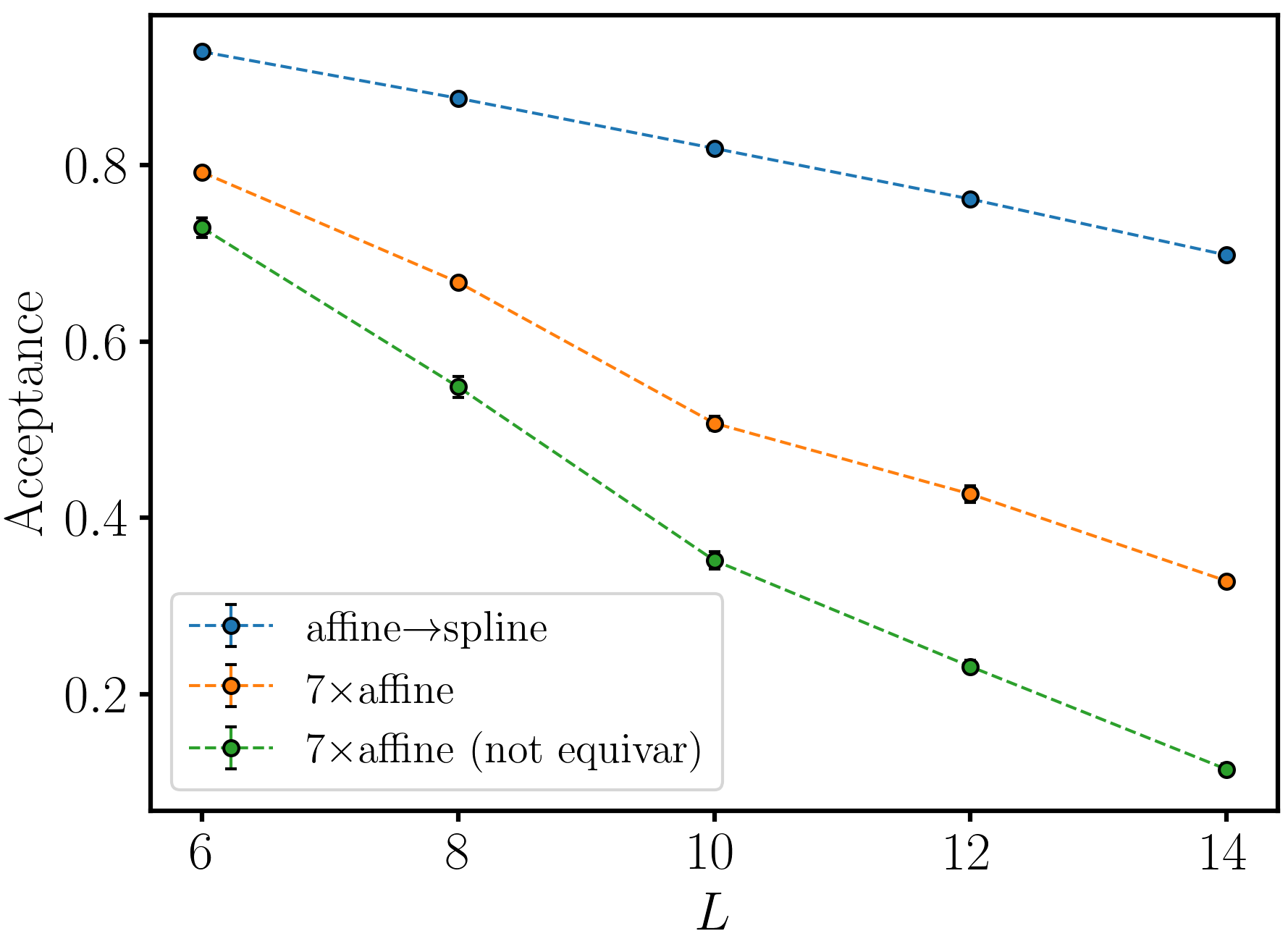}
        \caption{Comparison between three groups of models, each using approximately the same number of parameters in total; one pair of spline layers contains approximately the same number as six pairs of affine layers, given that the networks have one hidden layer of a common size. In the legend, `affine' and `spline' refer to blocks of two coupling layers that together transform all of the field variables once.}
        \label{fig:affine_vs_spline}
    \end{subfigure}
    ~
    \begin{subfigure}{.49\textwidth}
        \includegraphics[width=\textwidth]{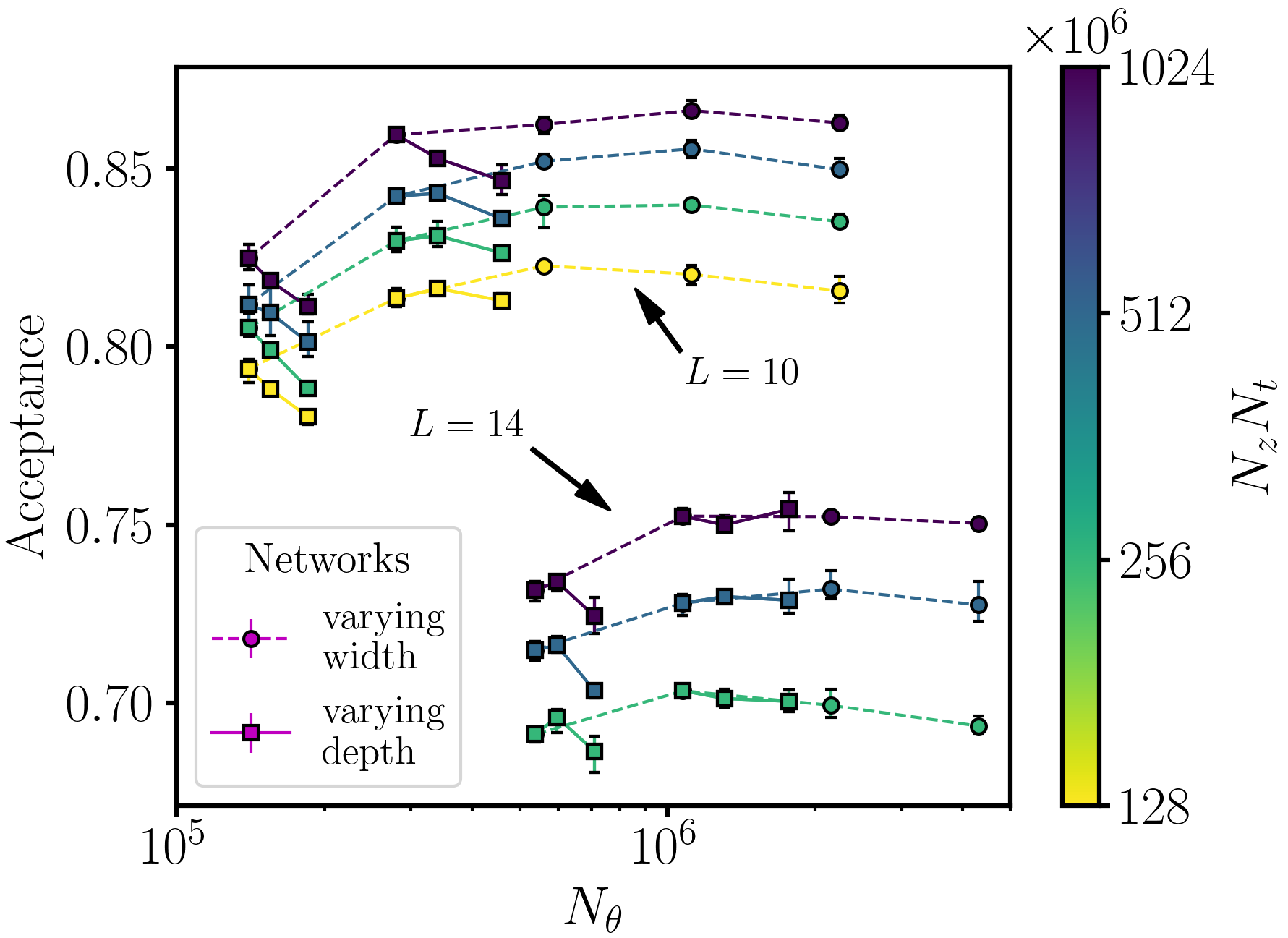}
        \caption{Acceptance rates resulting from varying the width (follow dotted lines) and depth (follow filled lines) of the neural networks within coupling layers. The $x$-axis represents the number of trainable parameters, $N_\theta$, in the entire flow, and the colour axis measures the extent of the training, $N_t$ being the number of steps involving batches of $N_z$ training examples.}
        \label{fig:params_vs_configs}
    \end{subfigure}
    \caption{}
\end{figure}

The original proof-of-principle study~\cite{Albergo2019} used normalizing flows built out of \textit{affine} coupling layers~\cite{Dinh2016} parametrised by deep fully-connected neural networks.
As we were experimenting, we noticed that three key modifications led to substantial improvements in acceptances and autocorrelations (see also Figure~\ref{fig:affine_vs_spline}):
\begin{enumerate}
    \item Upgrade the final pair of affine layers to more flexible \textit{rational quadratic spline} layers~\cite{Durkan2019b}.
    \item Enforce \textit{equivariance} with respect to the $\mathbb{Z}_2$ symmetry in the preceding affine layers.
    \item Replace the deep neural networks with shallow ones --- a single hidden layer suffices.
\end{enumerate}
By computing the two-point correlation function, and hence estimating the correlation length, at various stages in the transformation (i.e. on the intermediate states, $v_i$), we concluded that the initial affine layers mainly resolved strong correlations at short separations, whereas the final pair of spline layers split the unimodal distribution into a bimodal one (with modes at $\pm \langle \phi \rangle$), and resolved weaker correlations.
We also found it beneficial to forcefully anneal the learning rate from an initial maximum value down to zero at the end of training, allowing progressively weaker correlations to be learned via increasingly delicate optimisation steps.
Like Ref.~\cite{Hackett2021}, we found that annealing the temperature, $\beta^{-1}$, during training also slightly improved results, particularly in the strongly bimodal phase of $\phi^4$.
Presumably, both of these strategies result in `learning' being more uniformly distributed among the training iterations.

The key motivation behind the flow-based approach is that that once a normalizing flow model has been trained sampling is extremely efficient.
Given what appeared to be significant improvements in acceptance rates with respect to the proof-of-principle study~\cite{Albergo2019}, we were interested to determine how the cost of training these models could be expected to scale towards the continuum limit, since this `overhead' cost is ultimately the crucial factor that will determine whether this approach is more efficient in practice.

Before concentrating on the scaling we took some time to establish which hyper-parameters had the greatest influence over the quality of the trained model.
The key conclusions from these tests, exemplified by Figure~\ref{fig:params_vs_configs}, were that increasing the size of models (bigger neural networks, more layers, etc.) had a small and sometimes adverse effect on the sampling efficiency of the trained model, whereas acceptances were strongly dependent on two training hyper-parameters: the batch size ($N_z$ from earlier) and the number of training iterations, $N_t$.
We also did a post-hoc check that our results were not unduly poor due to the use of fully-connected neural networks ($\mathbf{N}$) in the coupling layers rather than convolutional neural networks.
In fact what we observed was that models using fully-connected networks reached higher acceptance rates than those using convolutional networks, with far less time spent training.
There are, however, several reasons to expect this hierarchy to invert as lattices and flow models increase in size, and we intend to investigate this in a future study.

We conducted a first investigation into the scaling; we restricted ourselves to effectively one, carefully chosen architecture involving up to five pairs of affine coupling layers followed by a final pair of spline coupling layers, each parametrised by a depth-two fully-connected neural network with $L^2$ neurons in the hidden layer.
We fixed the correlation length at $\xi \approx L/4$ and increased the lattice size from $L^2 = 6^2, 8^2, \ldots, 20^2$.
The largest models had $N_\theta<10^7$ trainable parameters, and the most extensive training schedule involved $N_t=32,000$ training updates, each involving a batch of $N_z=32,000$ configurations.

\begin{figure}
\centering
    \begin{subfigure}{.49\textwidth}
        \includegraphics[width=\textwidth]{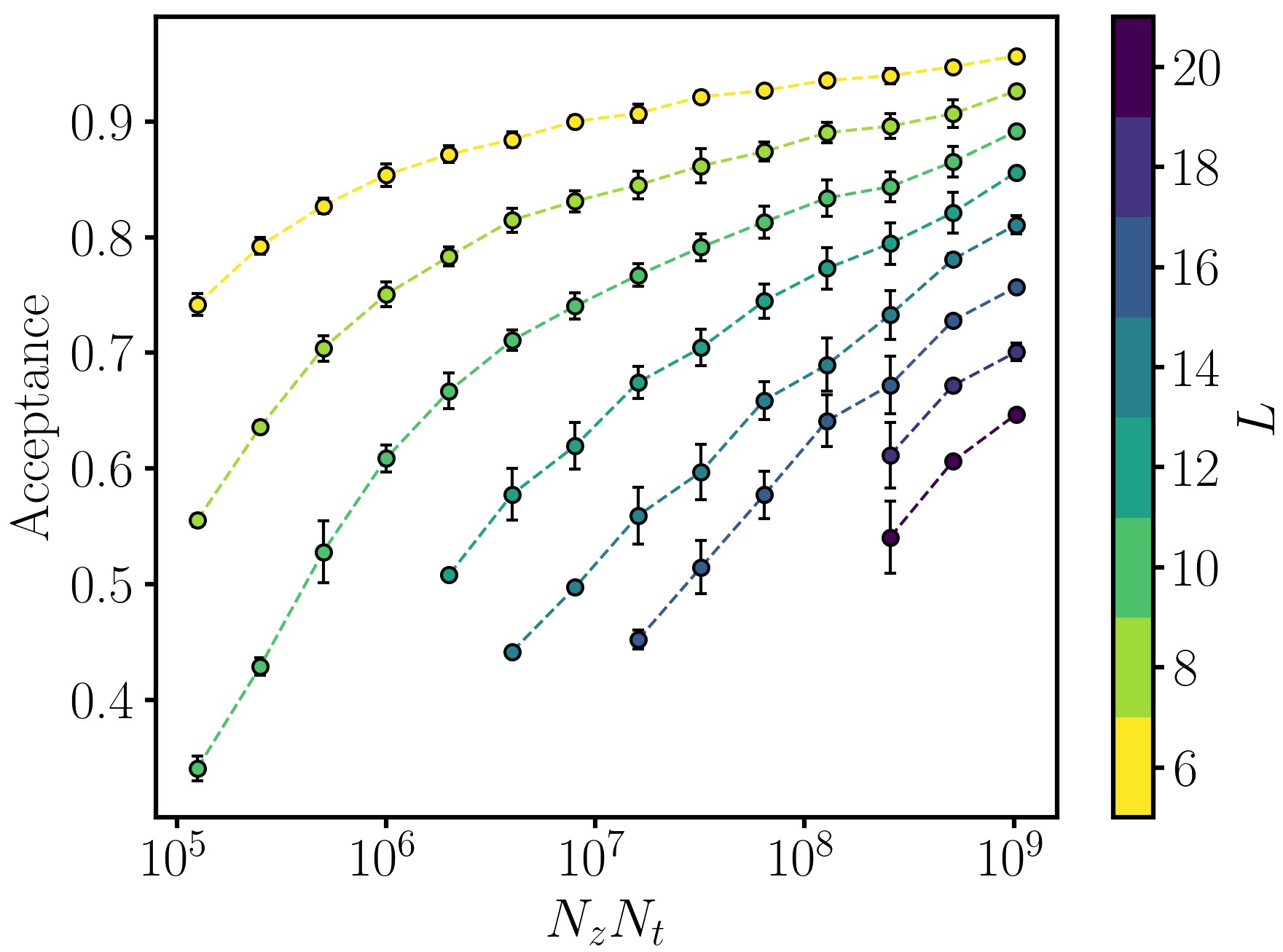}
        \caption{
        Average acceptance rates from fully-trained models, grouped by the number $N_z N_t$ of training examples (field configurations) from which they were able to learn. Data points and error bars are means and standard deviations taken over multiple, independently trained models, with different ratios $N_t/N_z =1, 2, 4$, and different numbers of affine layers.}
        \label{fig:configs_vs_acceptance}
    \end{subfigure}
    ~
    \begin{subfigure}{.49\textwidth}
        \includegraphics[width=\textwidth]{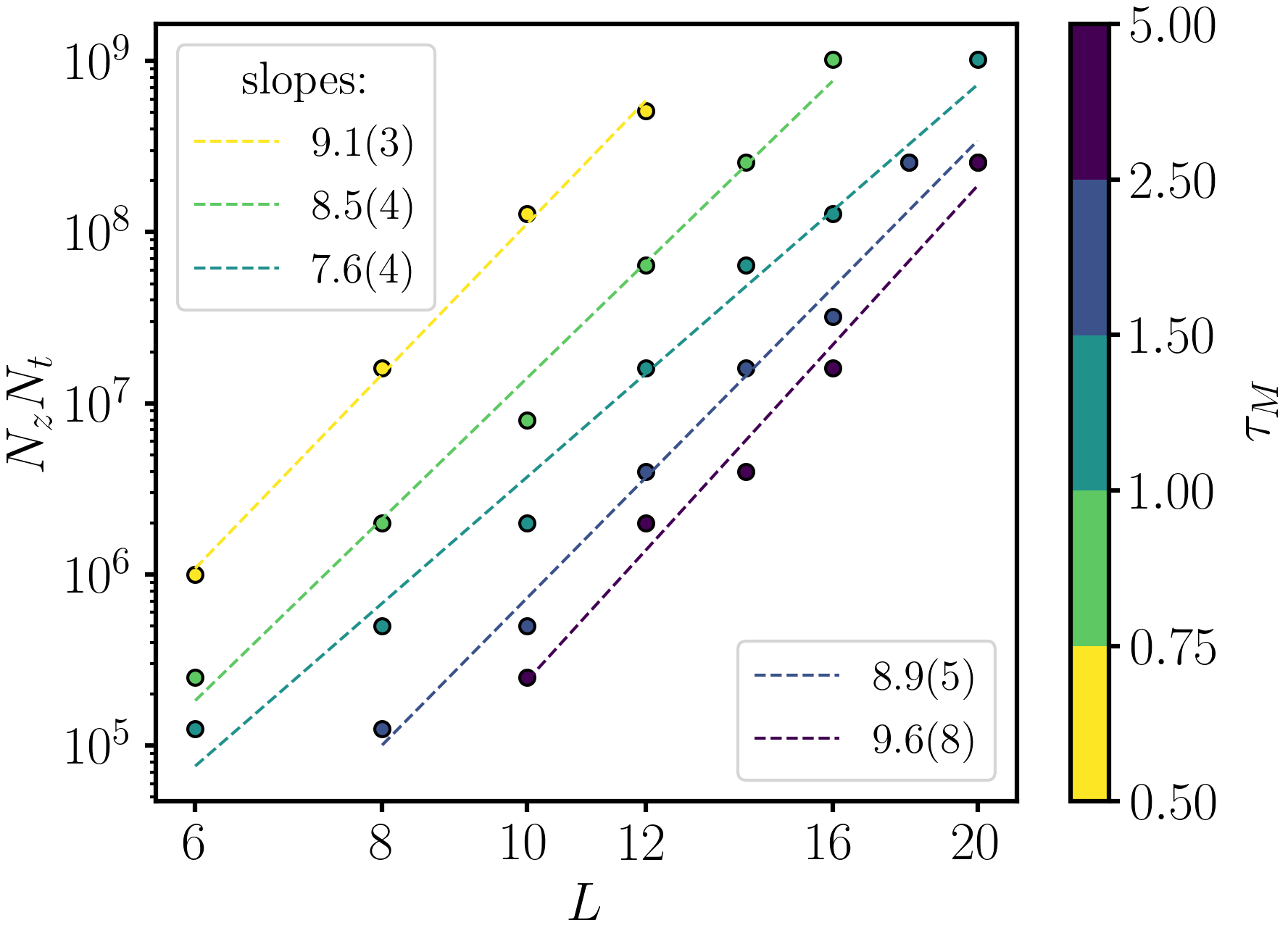}
        \caption{
        Scaling of `training costs', defined as the number $N_z N_t$ of training examples that were required to be generated. Models were grouped by integrated autocorrelation time, and the model with the lowest $N_z N_t$ selected from each group. This particular architecture appears to be very far from being competitive with traditional local-updates MCMC, where costs typically scale as $\sim L^{d+2}$ ($z_\mathcal{O}\approx 2$).}
        \label{fig:scaling}
    \end{subfigure}
    \caption{}
\end{figure}

More so than $N_z$ and $N_t$ individually, results were sensitive to the product $N_z N_t$, i.e. the number of `training examples' seen by the model.
Surprisingly, we did not observe any plateauing of acceptances (on a log scale) as $N_z N_t$ increased over 3 orders of magnitude, even for the smallest lattice size --- see Figure~\ref{fig:configs_vs_acceptance}.
This suggests that the limiting factor is not the inherent capacity of the network to approximate a trivializing map (i.e. expressivity) but begs the question: what is the origin of this slow convergence, and of the astonishingly poor scaling of training costs depicted in Figure~\ref{fig:scaling}?

A first, uncontroversial interpretation of slow convergence might be that the latter stages of optimisation require a large number (large $N_t$) of very precise (large $N_z$) steps.
However, this statement does not have any explanatory power.
Nor does it provide any insight into why the error bars in Figure~\ref{fig:configs_vs_acceptance} are so small; i.e. why the acceptance rate depends strongly on the product $N_z N_t$, but far less so on $N_z$ or $N_t$ individually, or on the number of layers/parameters in the model.

Drawing on observations made in Ref.~\cite{Huang2018b}, we suggest a candidate explanation, that is, after an initial training phase where the acceptance grows from effectively zero to the order of $10^{-1}$ rather quickly, the rate at which \textit{useful} training examples are generated is extremely low.
By `useful' we mean that the training example produces a contribution to the training update (a gradient) that causes the density $\tilde{p}(\phi)$ to expand along a specific direction in which it is currently underestimating $p(\phi)$.
To justify this we begin by stating two well-established facts.
Firstly, the typical behaviour of a training algorithm based on a loss function derived from Eq.~\eqref{eq:kullbach_leibler} is to quickly pick out the mode(s) of $p(\phi)$.\footnote{With the major caveat that it is possible to completely miss a subset of the modes.}
Secondly, as we increase the number of degrees of freedom in the system or increase the correlation length, $p(\phi)$ becomes increasingly concentrated on a low-dimensional manifold embedded in $\mathbb{R}^{L^2}$.
Recalling that $\tilde{p}(\phi)$ is an isotropic Gaussian at $t=0$, this suggests an initial, fast `mode-seeking' phase, after which $\tilde{p}(\phi)$ is likely to overestimate $p(\phi)$ in almost every direction.
Since training examples are generated with a probability proportional to $\tilde{p}(\phi)$, it follows that a large proportion of a training update will be devoted to attempting to compress $\tilde{p}(\phi)$ ever closer to the nearest mode of $p(\phi)$, but many of these gradients will oppose each other because total probability mass is conserved.
Consequently, the optimisation becomes heavily dependent on those relatively few training examples which produce gradients that guide the expansion of $\tilde{p}(\phi)$ along the manifold, away from the nearest mode.
Assuming this expansion phase is slow, the total number of these useful configurations that is generated during training grows in proportion with $N_z N_t$, and is of course oblivious to any efforts to improve results by changing the model itself.

In more general terms, given a highly correlated $p(\phi)$, there are theoretical grounds to expect a training scheme based on minimising Eq.~\eqref{eq:loss_fn} to encounter this very inefficient expansion phase~\cite{Huang2018b}.
However, while this provides a strong motivation to consider this as a potential contributing factor to the terrible scaling of our models, we wish to be clear that this is a \textit{candidate} explanation that needs verification.
For example, it is possible that this issue would become severe on large lattices, but is actually relatively mild on the small lattices we studied.
What is required is a far more systematic investigation aimed at disentangling the various factors contributing to the scaling of training costs.
As a next step, it would be instructive to separately quantify the effect of increasing the number of degrees of freedom and increasing the correlation length on training costs.
Once the picture has become clearer, if our description of a slow expansion phase turns out to be accurate, we can begin to try to alleviate the problem by, for example, modifying the training scheme.


\section{Conclusions}

We conducted a short study of the scaling of training costs based on an improved formulation of the normalizing flow models presented in Ref.~\cite{Albergo2019}.
Our results show that the particular combination of model architecture and training scheme used in several studies including our own~\cite{Albergo2019,Nicoli2020,DelDebbio2021} does not scale up in a way that would make it competitive with traditional MCMC sampling techniques.
It would be very unwise, however, to extrapolate from these conclusions; it is not yet clear what the root cause of the rapidly increasing training costs is, therefore neither is it clear whether this is a problem that can be alleviated or circumvented.
Our intention is to scale up this study to much larger lattices and systematically quantify the factors influencing the scaling of training costs by varying the model architecture (including using convolutional networks), training algorithm, and hyper-parameters.
We also aim to incorporate other lattice field theories, such as the topologically non-trivial $\mathrm{CP}^{N-1}$ models, into this study.

\newpage
\bibliographystyle{JHEP}
\bibliography{bibtex.bib}

\providecommand{\href}[2]{#2}\begingroup\raggedright\begin{thebibliography}{10}

\bibitem{Duane1987}
S.~Duane, A.D.~Kennedy, B.J.~Pendleton and D.~Roweth, \emph{Hybrid {Monte
  Carlo}}, \href{https://doi.org/10.1016/0370-2693(87)91197-x}{\emph{Phys.
  Lett. B} {\bfseries 195} (1987) 216}.

\bibitem{Luscher2009}
M.~L\"{u}scher, \emph{Trivializing maps, the {Wilson} flow and the {HMC}
  algorithm}, \href{https://doi.org/10.1007/s00220-009-0953-7}{\emph{Commun
  Math Phys} {\bfseries 293} (2009) }
  [\href{https://arxiv.org/abs/0907.5491}{{\ttfamily 0907.5491}}].

\bibitem{Albergo2019}
M.S.~Albergo, G.~Kanwar and P.E.~Shanahan, \emph{Flow-based generative models
  for {Markov chain Monte Carlo} in lattice field theory},
  \href{https://doi.org/10.1103/physrevd.100.034515}{\emph{Phys. Rev. D}
  {\bfseries 100} (2019) 034515}
  [\href{https://arxiv.org/abs/1904.12072}{{\ttfamily 1904.12072}}].

\bibitem{Kanwar2020}
G.~Kanwar, M.S.~Albergo, D.~Boyda, K.~Cranmer, D.C.~Hackett, S.~Racani\`{e}re
  et~al., \emph{Equivariant flow-based sampling for lattice gauge theory},
  \href{https://doi.org/10.1103/physrevlett.125.121601}{\emph{Phys. Rev. Lett.}
  {\bfseries 125} (2020) 121601}
  [\href{https://arxiv.org/abs/2003.06413}{{\ttfamily 2003.06413}}].

\bibitem{Boyda2020}
D.~Boyda, G.~Kanwar, Racani\`{e}re, D.J.~Rezende, M.S.~Albergo, K.~Cranmer
  et~al., \emph{Sampling using {$\mathrm{SU}(N)$} gauge equivariant flows},
  \href{https://doi.org/10.1103/PhysRevD.103.074504}{\emph{Phys. Rev. D}
  {\bfseries 103} (2021) 074504}
  [\href{https://arxiv.org/abs/2008.05456}{{\ttfamily 2008.05456}}].

\bibitem{Albergo2021b}
M.S.~Albergo, G.~Kanwar, S.~Racani\`ere, D.J.~Rezende, J.M.~Urban, D.~Boyda
  et~al., \emph{Flow-based sampling for fermionic lattice field theories},
  \href{https://arxiv.org/abs/2106.05934}{{\ttfamily 2106.05934}}.

\bibitem{Hackett2021}
D.C.~Hackett, C.-C.~Hsieh, M.S.~Albergo, D.~Boyda, J.-W.~Chen, K.-F.~Chen
  et~al., \emph{Flow-based sampling for multimodal distributions in lattice
  field theory},  \href{https://arxiv.org/abs/2107.00734}{{\ttfamily
  2107.00734}}.

\bibitem{Dinh2016}
L.~Dinh, J.~Sohl-Dickstein and S.~Bengio, \emph{Density estimation using {Real
  NVP}},  \href{https://arxiv.org/abs/1605.08803}{{\ttfamily 1605.08803}}.

\bibitem{Nicoli2020}
K.A.~Nicoli, C.J.~Anders, L.~Funcke, T.~Hartung, K.~Jansen, P.~Kessel et~al.,
  \emph{On estimation of thermodynamic observables in lattice field theories
  with deep generative models},
  \href{https://doi.org/10.1103/physrevlett.126.032001}{\emph{Phys. Rev. Lett.}
  {\bfseries 126} (2021) 032001}
  [\href{https://arxiv.org/abs/2007.07115}{{\ttfamily 2007.07115}}].

\bibitem{DelDebbio2021}
L.~Del~Debbio, J.~Marsh~Rossney and M.~Wilson, \emph{Efficient modelling of
  trivializing maps for lattice {$\phi^4$} theory using normalizing flows: A
  first look at scalability},
  \href{https://doi.org/https://doi.org/10.1103/PhysRevD.104.094507}{\emph{Phys.
  Rev. D} {\bfseries 104} (2021) 094507}
  [\href{https://arxiv.org/abs/2105.12481}{{\ttfamily 2105.12481}}].

\bibitem{Durkan2019b}
C.~Durkan, A.~Bekasov, I.~Murray and G.~Papamakarios, \emph{Neural spline
  flows}, {\emph{Advances in Neural Information Processing Systems} {\bfseries
  32} (2019) } [\href{https://arxiv.org/abs/1906.04032}{{\ttfamily
  1906.04032}}].

\bibitem{Huang2018b}
C.-W.~Huang, F.~Ahmed, K.~Kumar, A.~Lacoste and A.~Courville, \emph{Probability
  distillation: A caveat and alternatives}, {\emph{Proceedings of The 35th
  Uncertainty in Artificial Intelligence Conference} {\bfseries 115} (2020)
  1212}.

\end{thebibliography}\endgroup

\acknowledgments

LDD is supported by an STFC Consolidated Grant, ST/P0000630/1, and a Royal Society Wolfson Research Merit Award, WM140078.
JMR is supported by STFC, grant ST/T506060/1.
MW is supported by STFC, grant ST/R504737/1.
This work has made use of the resources provided by the Edinburgh Compute and Data Facility (ECDF).

\end{document}